\documentclass[
aps,%
12pt,%
final,%
notitlepage,%
oneside,%
onecolumn,%
nobibnotes,%
nofootinbib,
superscriptaddress,%
noshowpacs]%
{revtex4}
\usepackage{epsfig}

\begin{document}
\title{Elliptic Flow in Central Collisions of Deformed Nuclei}
\author{\firstname{Peter} \surname{Filip}}
\email[]{fyziflip@savba.sk}
\affiliation{Institute of Physics, Slovak Academy of Sciences, 
Bratislava, Slovakia }
\noaffiliation
\begin{abstract}
Non-trivial geometrical effects in relativistic central collisions
of deformed nuclei are studied using a simple version of optical Glauber model.
For very small impact parameters
large centrality and eccentricity fluctuations are observed. 
In very high multiplicity collisions of oblate nuclei 
(e.g. Au$^{197}$ and Cu) a significant fraction of events with 
elliptic flow strength $v_2$ dependent on oblateness $\beta_2$ is predicted.
\end{abstract}
\maketitle
\section{Introduction.}
In relativistic heavy-ion collisions a simple
relation between {\it centrality\,} 
of collisions $c(N)$ and average participant {\it eccentricity}
of interacting nucleons $< \epsilon _p >$
is usually expected to be valid. According to rather general assumptions
\cite{WBronWFlor} centrality $c(N)$
is directly related to impact parameter
$b$ as $c(N) \approx {\pi b^2}/{\sigma_{inel}}$.
At the same time the participant eccentricity 
\begin{equation}
\epsilon _p (b) = {\sum_i (x^2_i - y^2_i)}/{\sum_i (x^2_i + y^2_i)}
\end{equation}
(where $x_i$ and $y_i$ are positions of interacting nucleons 
in transversal plane) is also singly dependent (within fluctuations)
on the size of impact parameter $b$. Thus
measured elliptic flow strength $v_2$  
is expected to rise \cite{JYOPRD,JYOPLB} from very small values in  
most central collisions proportionaly to 
increasing values of average eccentricity 
$v_2 \approx k \cdot < \epsilon _p(b) >$. 

However these simple relations are not valid for collisions of
deformed nuclei. The orientation
of a deformed nucleus (e.g. relative to beam axis) has a direct influence on
centrality and eccentricity of collisions 
at a given fixed value of impact parameter $b$.
This is slightly worrying since the elliptic flow strength measured in 
collisions
of nuclei having non-zero deformation $\beta_2$ (e.g. Au, Cu) has 
been interpreted \cite{RHICv2}
assuming these nuclei to be spherical.

In the following sections we present a preliminary study of
non-trivial variations of initial eccentricity
at given collision centralities in relativistic collisions 
of deformed nuclei.


\section{Geometrical Effects in Collisions of Deformed Nuclei}
Although effects of nuclear deformation have been 
carefully studied in fusion reactions at coulomb barrier \cite{CF}
the role of nuclear deformation in relativistic heavy ion collisions have been
considered scarcely so far. It had been pointed out 
\cite{Shuryak}
that highest energy
densities of QCD matter could be created in central collisions 
of longitudinaly oriented heavy deformed nuclei and elliptic flow in such collisions
had been discussed. 
However since the heaviest nuclei suitable for relativistic collisions
experiments are prolate \cite{MollerB2} the 
elliptic flow in central collisions of oblate nuclei have escaped
the attention of heavy ion community so far.
 
In this contribution we point out that oblate ground-state nuclear 
deformation of stable nuclei can have 
a significant influence on measured elliptic flow values in central collisions
since it directly influences fluctuations and average
values of initial eccentricity $\epsilon _p$ at given collision centralities. 

\begin{figure}
\includegraphics[width=12cm]{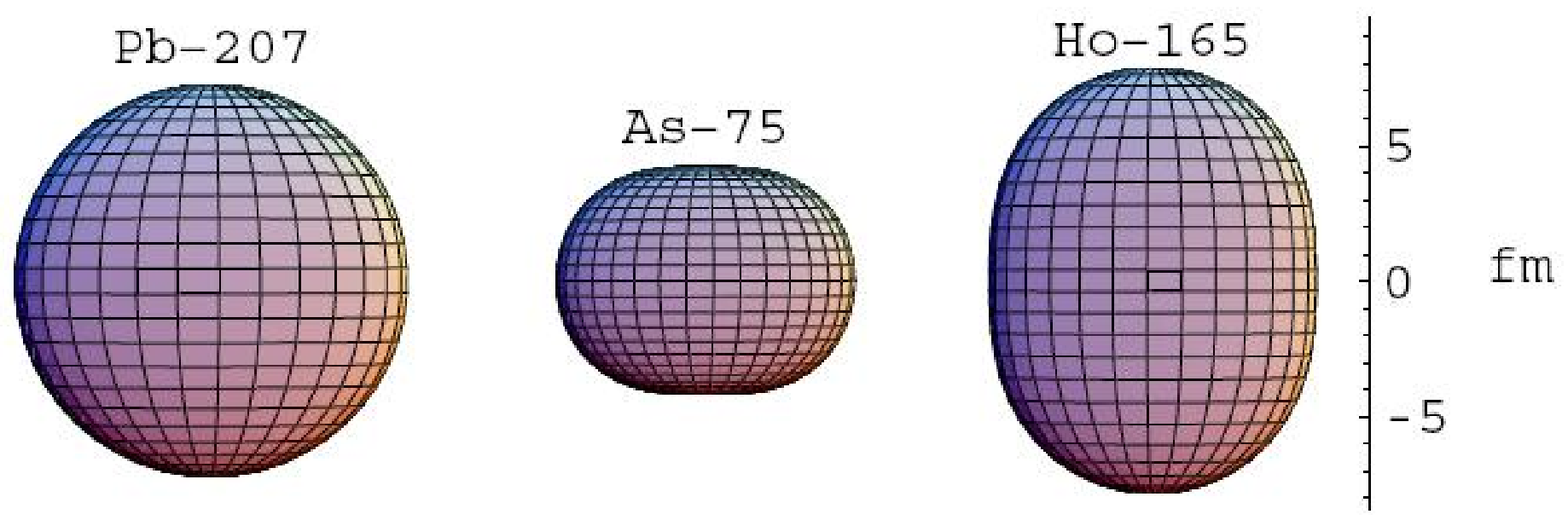}

{\small
{\bf Fig.1:} Predicted shapes of Pb$^{207}$, As$^{75}$ ($\beta_2=-0.25$)
and Ho$^{165}$ ($\beta_2=0.3$) nuclei \cite{MollerB2}. 
}
\label{fig:Shapes1}
\end{figure}

Let us consider angle $\theta $ between the beam direction and the main axis
of a deformed nucleus (e.g. Ho-165). 
For $\theta \approx 0$ nucleus is longitudilay polarized
and for $\theta \approx \pi/2$ we have transversal polarization.
Even for very small impact parameters $b\approx 0$
(central collisions geometrically)
the orientation of a deformed nucleus relative to beam axis
influences directly participant eccentricity
\begin{equation}
\epsilon _p (\theta, b) = \frac{\sum^i x^2_i - y^2_i}{\sum^i x^2_i + y^2_i}
\end{equation}
and thus final strength of the elliptic flow observed.
Also {\it centrality} of collision
depends on the orientation of a deformed nucleus at given impact parameter.
This happens not only due to changes in the number of participating nucleons
(a simple overlap effect) but also due to that fraction of secondary particles 
multiplicity which is proportional to total number of binary nucleon-nucleon 
collisions \cite{DimaK}.

Thus geometric relation between centrality and impact parameter
\cite{WBronWFlor}
is not valid for collisions of deformed nuclei and initial
eccentricity $\epsilon_p$ is not simply related to collision {\it centrality}
as it is usually assumed. This can in principle modify interpretations drawn from the
eliptic flow strength $v_2$ at given centrality $c(N)$ 
in Au+Au collisions at RHIC.

\section{A SIMPLE OPTICAL GLAUBER SIMULATION}
In order to test quantitatively the effects of nuclear deformation
on centrality and eccentricity in heavy ion collisions
a simple Optical Galuber Model (OGM) calculation for deformed nuclei
had been performed. Here is a very short description of the simulation:

In the first step a deformed Woods-Saxon \cite{WoodsSaxD}
density
\begin{equation}
\rho (\vec r) = \frac{\rho_o}{1+e^{(r-R_0(1+\beta_2 Y_{20}+\beta_4 Y_{40}))/a}}
\end{equation}
with deformation parameters $\beta _2, \beta _4$
taken from \cite{MollerB2}
for various nuclei (Au,Cu,Ho,Pb,Si,Ca)
had been used to generate
transversal projections $\rho^T(x,y) = \int \rho (\vec r) dz$ 
of nucleon density.
Binning of $\rho ^T(x_i,y_j)$ histograms was $0.1${\it fm}
and longitudinal integration of deformed
Woods-Saxon density $\rho(\vec r)$ was performed in 
0.05{\it fm} steps.
For deformed nuclei separate projections $\rho^T[\theta,\phi]$ had been
generated at fixed values of angle $\theta $ and azimuthal angle $\phi $.

\begin{figure}[h]
\includegraphics[width=16cm]{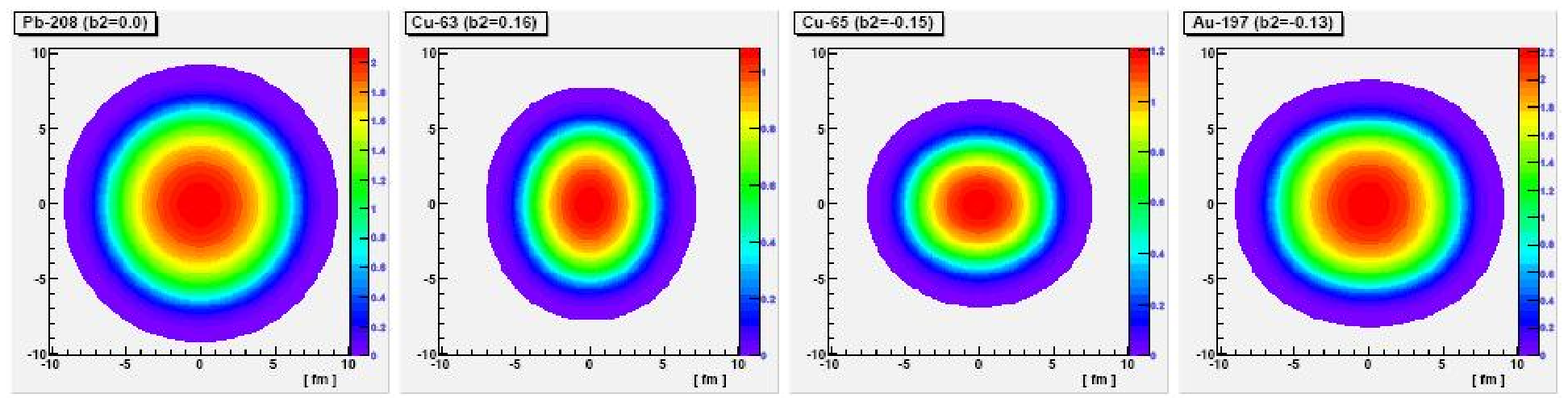}

{\small
{\bf Fig.2:} Projected nucleon densities $\rho ^T(x,y)$
for Pb$^{208}$, $^\uparrow$Cu$^{63}$, $^\uparrow$Cu$^{65}$, 
$^\uparrow$Au$^{197}$ 
($\beta _2$ from \cite{MollerB2}).
}
\label{fig:Shapes2}
\end{figure}

In order to simulate also non-central collisions 
transversal projections $\rho^T(x,y)$ had been shifted for selected
values of impact parameter $\vec b = (b_x,b_y)$ and 
total number of 
nucleon-nucleon collisions $N_{coll.}$ had been evaluated as
\begin{equation}
N_{coll}=\sum_{i,j}N_{coll}(x_i,y_j)=\sum_{i,j} \sigma^{inel}_{NN}\cdot  
                   \rho^T_1(x_i-b_x/2,y_j-b_y/2)* \rho^T_2(x_i+b_x/2,y_j+b_y/2).
\end{equation}
The cross section $\sigma^{inel}_{NN}$ was fixed at $42mb$.

Using the geometrical information stored in histograms $\rho^T_1(x_i,y_i)$ and
$\rho^T_2(x_i,y_i)$ transversal projected baryon density
$\rho_B(x,y)$ and number of participants $N_{part}$
were simply evaluated and found to be in good agreement\footnote{
Spherical shape for
Au-197 with $R=6.38$fm and diffusivity $a=0.53$fm was used for this comparison. 
}
with results of Monte-Carlo
Glauber model simulation \cite{AuAuRHO} as shown in Figs.9-10 (Appendix).  

Eccentricity $\tilde \epsilon _{NN}$ of the interacting volume 
had been evaluated as 

\begin{equation}
\tilde \epsilon_{NN} =  
\frac{\sum_{i,j} N_{coll}(x_i,y_j)\cdot (x_{i}^2-y_{j}^2)}{
      \sum_{i,j} N_{coll}(x_i,y_j)\cdot (x_{i}^2+y_{j}^2)}
\end{equation}
in CMS coordinates of the overlaping zone. 
Numerical values of eccentricity 
$\tilde \epsilon _{NN}$ were found to be in reasonable
agreement with Au+Au eccentricities evaluated in \cite{JYOPLB}
(see Fig.11 in Appendix).
Selected results of these calculations for 
Au,Cu,Ho,Pb,Si nuclei are presented in the following sections.

\section{Collisions of Spherical + Deformed Nuclei}
For the sake of simplicity let us start with polarized 
collisions of a deformed nucleus e.g. a prolate  
Ho-165 ($\beta_2 \approx 0.3$) with spherical projectile
Pb-207.

\begin{figure}[h]
\includegraphics[width=7.50cm]{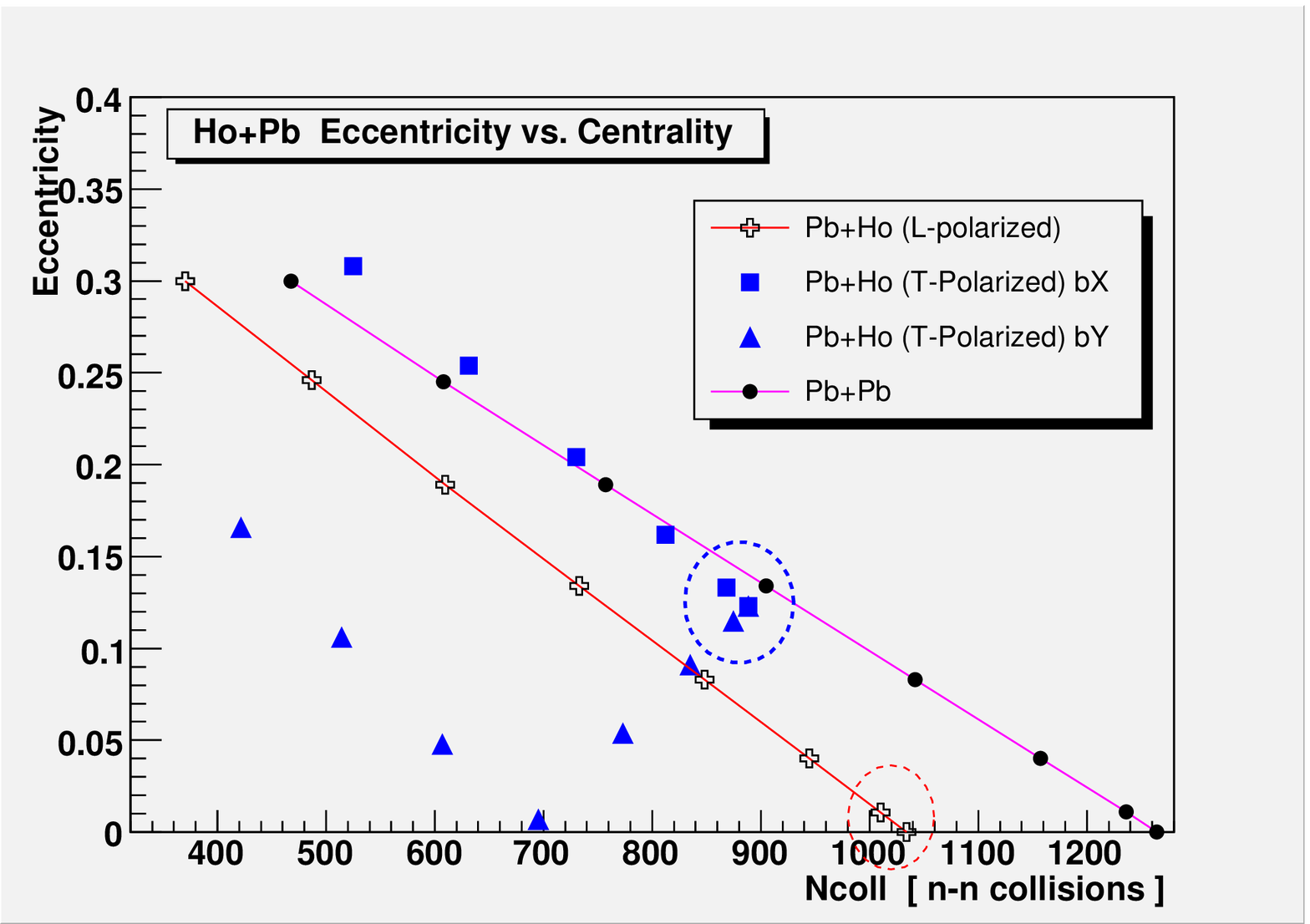}
\includegraphics[width=8.69cm]{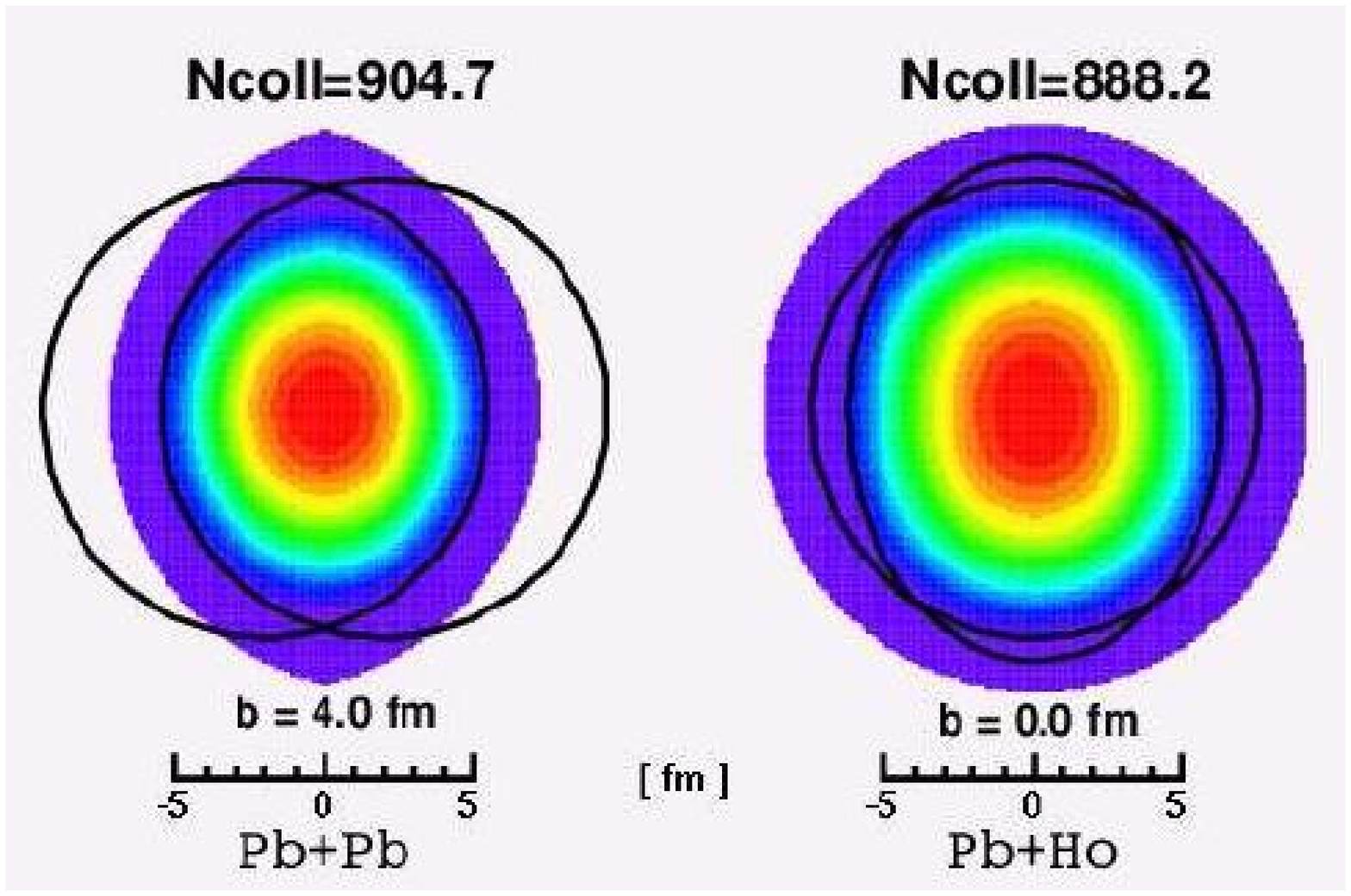}

{\small
{\bf Fig.3:} Eccentricity[n-n collisions] plot for Pb+Pb and 
polarized $^\uparrow$Ho+Pb systems.
}
\label{fig:Shapes3}
\end{figure}

If the spin$^{(7/2)}$ of Ho-165 nucleus
is orthogonal to beam axis (T-polarized) the participant
eccentricity $\epsilon_p$
will be non-zero even for $b=0$
its value roughly corresponding to
non-central Pb+Pb collisions at $b = 4$ fm (see Fig.3). Increasing
impact parameter values ($b_x,b_y$) in directions parallel $(b_y)$
and orthogonal $(b_x)$ 
to Ho$^{165}$ spin have different influence on the eccentricity. For 
increasing $(b_x,0)$ the eccentricity further increases while for
increasing $(0,b_y)$ the eccentricity first decreases to zero
at $b_y \approx 4$ fm and then it increases as it can be seen from  
Fig.3 (data points are evaluated for impact parameters increasing in steps 1fm).

For longitudinal polarization of holmium 
nuclei (relative to beam direction)
the initial eccentricity 
tends to zero in the most central collisions. This happens because
eccentricity fluctuations \cite{EccFluct} are not taken into account
in our simple optical Glauber calculation. It is also visible from
Fig.3 that central collisions of L-polarized Ho-165 with Pb
have significantly higher multiplicity compared to central T-polarized Ho+Pb
collisions (with $b=0$).
Thus the highest multiplicity collisions of prolate nuclei 
correspond to longitudinal orientations of deformed nucleus
due to the highest
number of nucleon-nucleon interactions in this case.

This means that in order to study the elliptic flow in central collisions
of transversaly oriented prolate nuclei with spherical
projectiles one needs to select events
with lower measured multiplicity then maximal. 
In such sample of events
non-central collisions of longitudinaly oriented prolate nuclei
will get mixed together with central collisions ($b \approx 0$) of transversaly
oriented prolate nuclei of similar multiplicity (see Fig.3).
One can try to distinguish these two types of events
via measurement of spectator energy in ZDC calorimeters and to perform
target (or beam) polarization-dependent experiments 
(see \cite{Ho156polar} for polarized Ho-165 target).  

\begin{figure}[h]
\includegraphics[width=7.9cm]{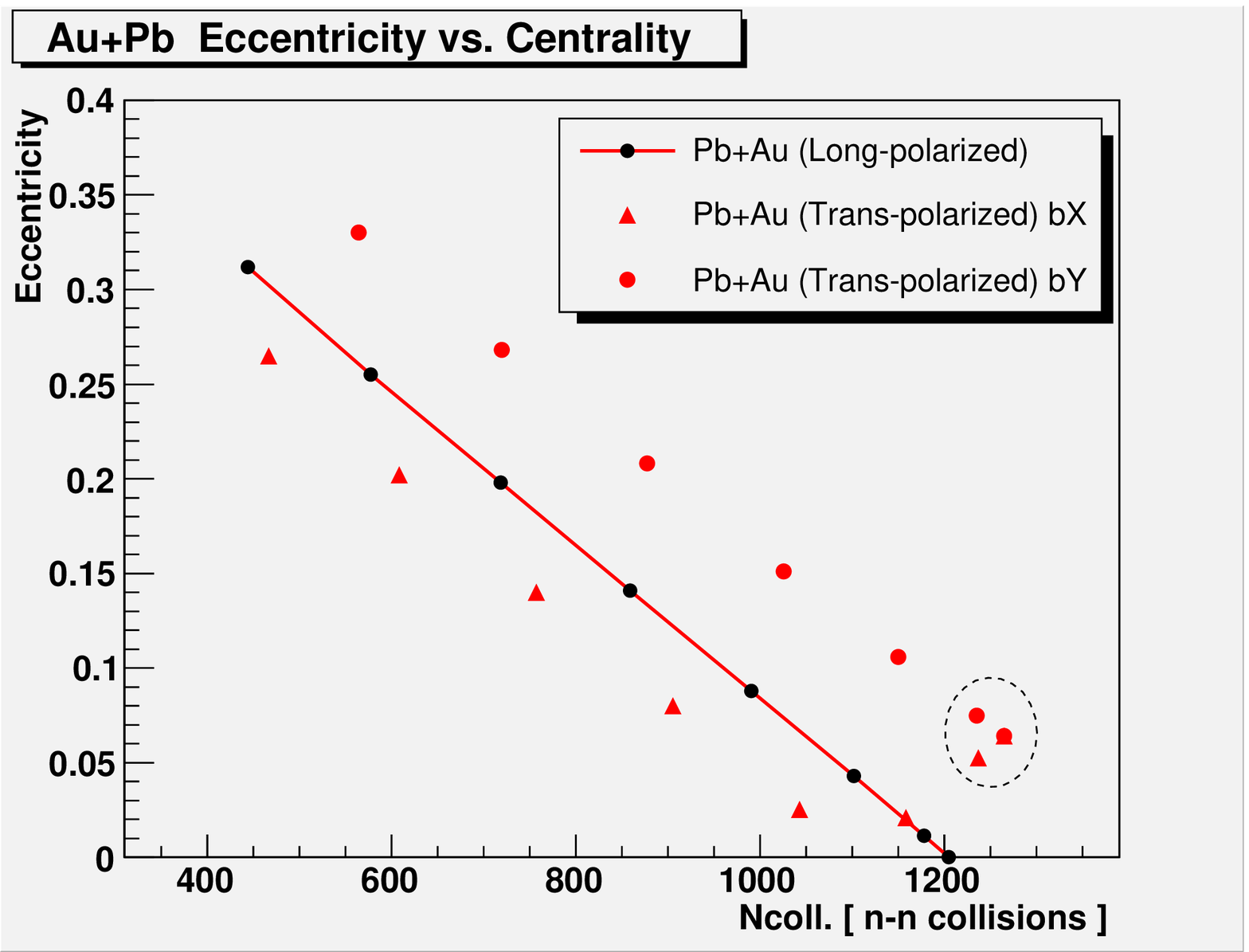}
\includegraphics[width=7.9cm]{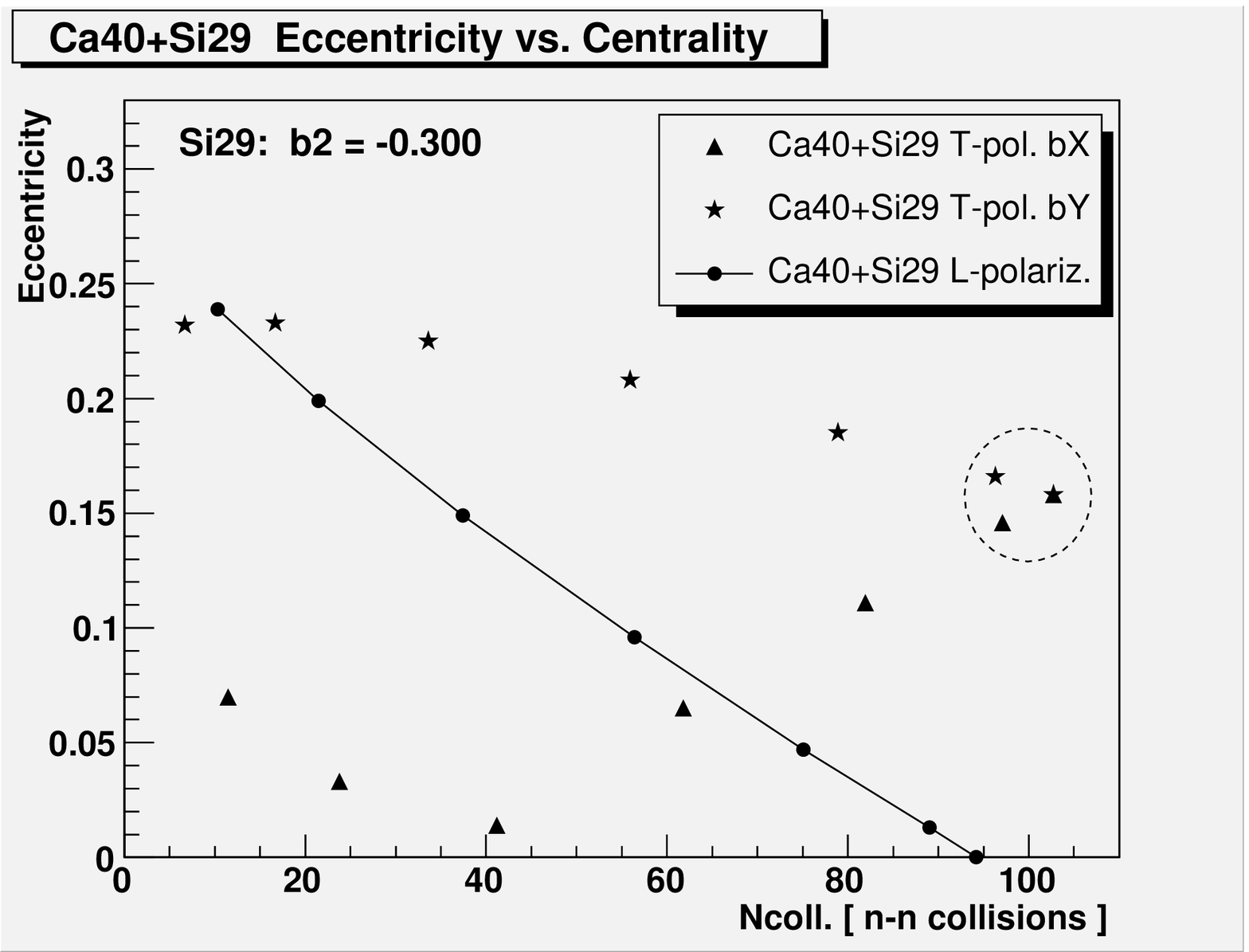}

{\small
{\bf Fig.4:} Eccentricity plots
 for polarized $^\uparrow$Au$^{197}$+Pb$^{207}$ and
$^\uparrow$Si$^{29}$+Ca$^{40}$ collisions. 
}
\label{fig:Shapes4}
\end{figure}

Fortunately
the existence of nuclei with oblate ($\beta_2 < 0$) deformation
(Au,I,Ga,As,Si,Al) provides us with another possibility in this direction.
For oblate nuclei 
very high multiplicity collisions correspond
to transversal polarization of nuclear spin relative to
the beam direction due to higher number of nucleon-nucleon 
collisions in this case. Thus significant fraction
of very central collisions of oblate nuclei should exhibit non-zero 
elliptic flow $v_2$ 
values proportional to deformation parameter $\beta_2 < 0$.
In Fig.4 we show results of our calculations for Au$^{197}$+Pb$^{207}$ and 
Si$^{29}$+Ca$^{40}$ collisions.
A small difference
in the number\footnote{One should always keep in mind the
existence of fluctuations in the number of n-n collisions.}
of nucleon-nucleon
interactions in the most central collisions with 
longitudial and transversal spin polarizations is obvious. 
In the next section we show that the influence of nuclear 
ground-state deformation on the eccentricity fluctuations
in Au+Au and Cu+Cu collisions (studied at RHIC) 
is expected to be more significant.

\section{Collisions of Two Deformed Nuclei}
In collisions of two deformed nuclei various geometrical
configurations are possible. Let us denote $\theta _1$ and 
$\theta _2 $
to be angles of deformed nuclei main axes relative to beam direction
and $\phi _1$ and $\phi _2$ to be main axes azimuthal angles in LAB. 
For unpolarized beams (targets)
azimuthal angles 
are independent and nuclei collide at random orientations
of $\phi _1, \phi _2 $. A trivial consideration shows
that probability of having a collision with 
deformed nuclei axes oriented at {\it relative} azimuthal angle 
$0 < \Delta \phi < \pi/2 $ is constant:
$
P(\Delta \phi ) = const$.

\begin{figure}[h]
\includegraphics[width=13cm]{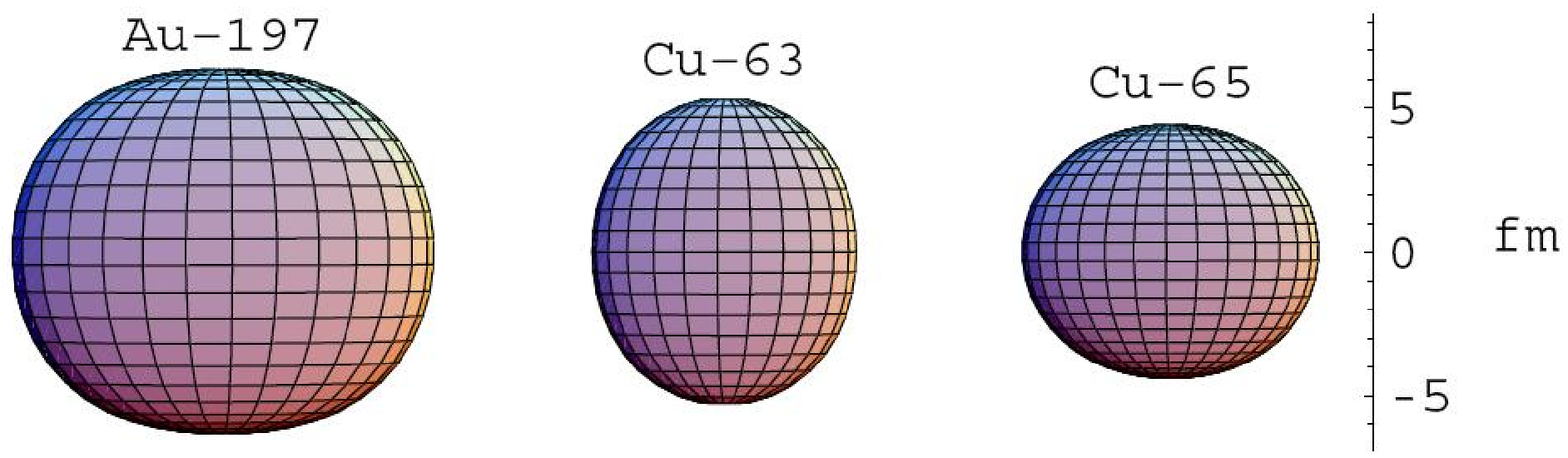}

{\small
{\bf Fig.5:} Shape of Au$^{197}$, Cu$^{63}$, Cu$^{65}$ with deformation 
$\beta_2=$[-0.13/+0.16/-0.15] taken from \cite{MollerB2}.
}
\label{fig:Shapes5}
\end{figure}

Polar angles $\theta_1, \theta _2$ are also mutually independent but 
not random and
probability $P(\theta_1,\theta_2)$ of a collision of two nuclei  oriented
at angles $0<\theta_1<\pi$ and $0<\theta _2<\pi$ is

\begin{equation}
P(\theta_1,\theta_2) = \sin (\theta_1) \sin (\theta_2) /4 
\end{equation}  
which says that longitudinal-longitudinal oriented (Long-Long) 
collisions 
of deformed nuclei are rather rare while transversal (Trans-Trans)
collisions ($\theta_1 \approx \theta_2 \approx \pi/2$)
are more frequent. 

One should keep in mind that this applies to collisions at
any value of geometrical impact parameter occuring with probability
$P(b<b_{max}) = c\cdot b_{max}^2$. Therefore central ($b\approx 0$) Long-Long collisions
are very rare! Nevetherless, just such collisions of {\it prolate} nuclei
can be selected \cite{Shuryak}
from the sample of events according to the centrality (high-multiplicity) criterion
and analyzed.

\begin{figure}[h]
\includegraphics[width=4.22cm]{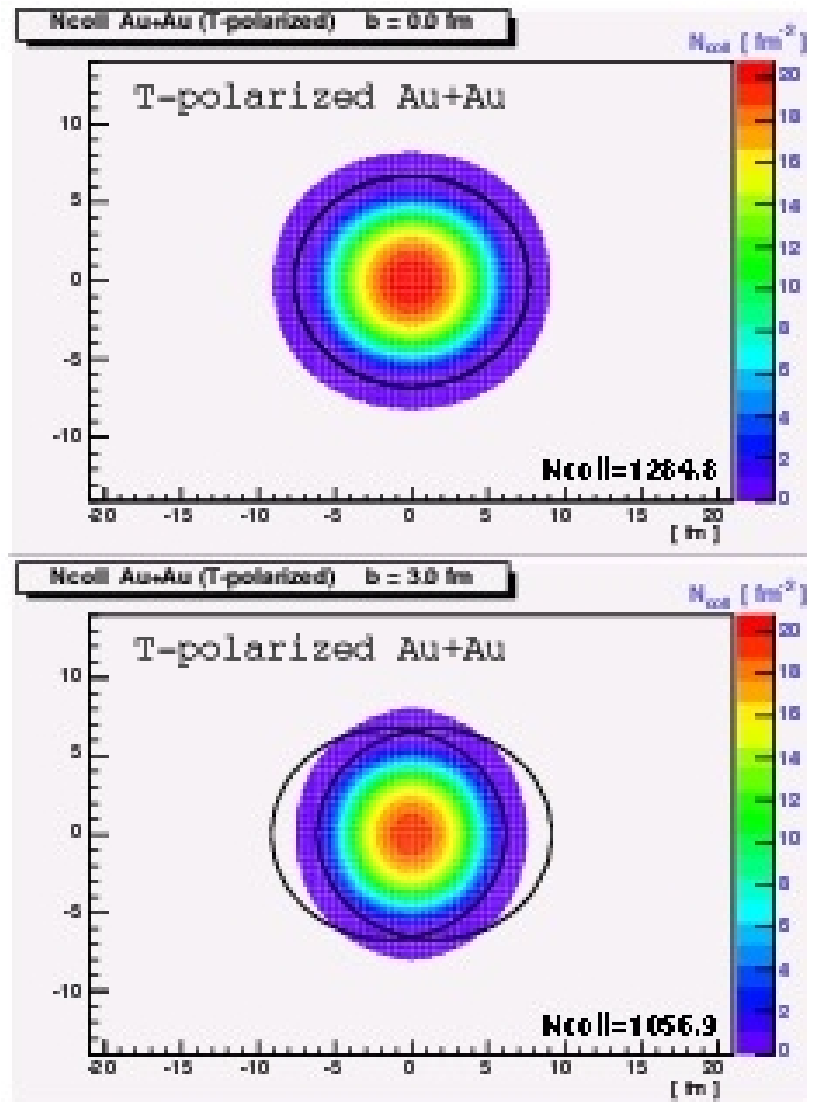}
\includegraphics[width=7.50cm]{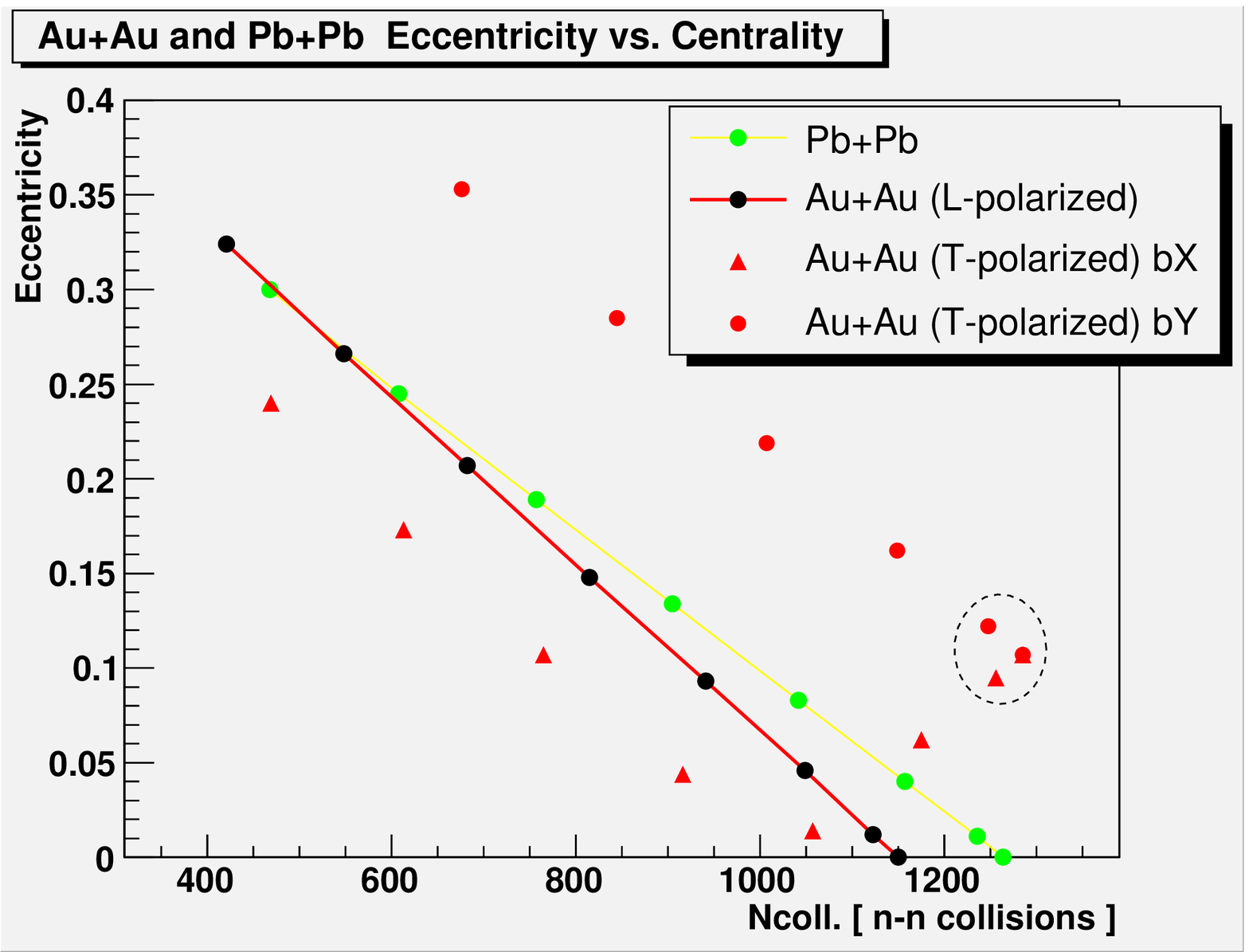}
\includegraphics[width=4.22cm]{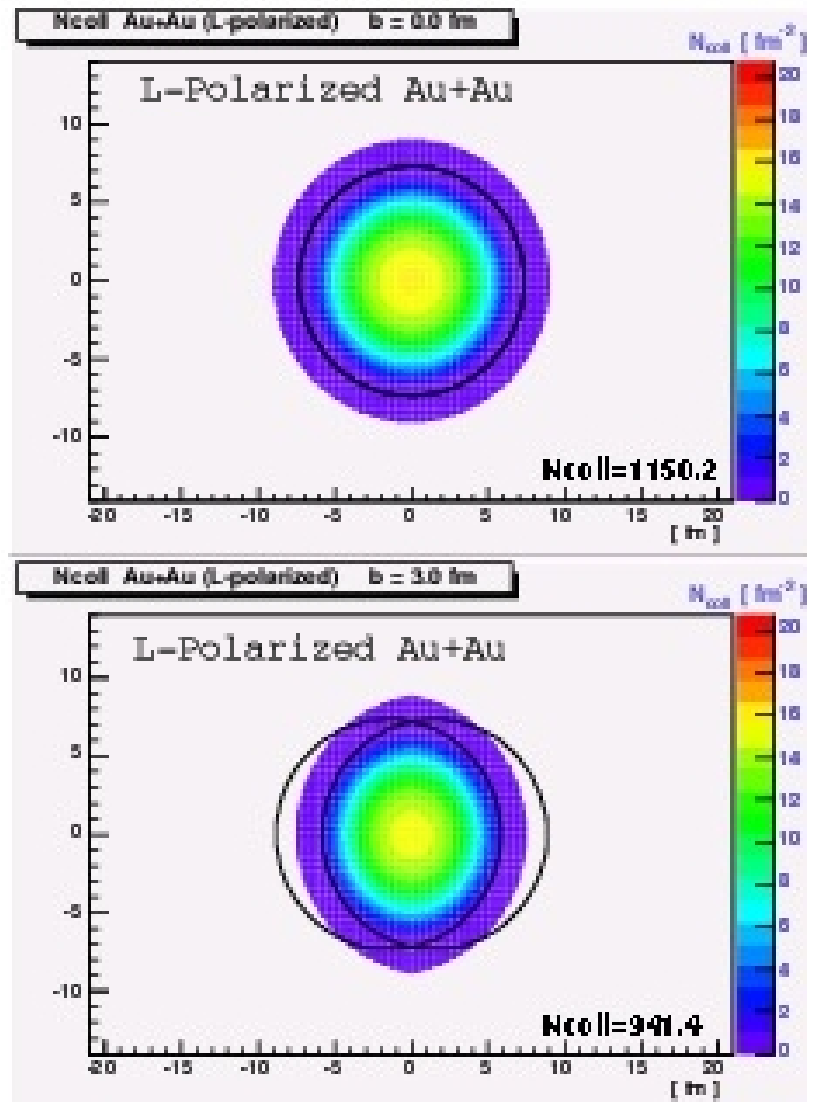}

{\small
{\bf Fig.6:} Densities $N_{coll}(x,y)/$fm$^{2}$ for 
Au$^{\uparrow}$+Au$^{\uparrow}$ at 
$b_x$=0fm/3fm and Eccentricity[Ncoll] plot. 
}
\label{fig:Shapes6}
\end{figure}

As can be seen from Fig.6 Trans-Trans collisions of {\it oblate} nuclei 
with parallel orientation of nuclear spins ($\Delta \phi = 0$)
are the highest multiplicity collisions for nuclei with oblate geometry.
They can be localized in the sample of events
based on the collision centrality and elliptic flow 
$v_2$ which is expected to be increasing with oblateness strength -$\beta_2$.

\begin{figure}[h]
\includegraphics[width=7cm]{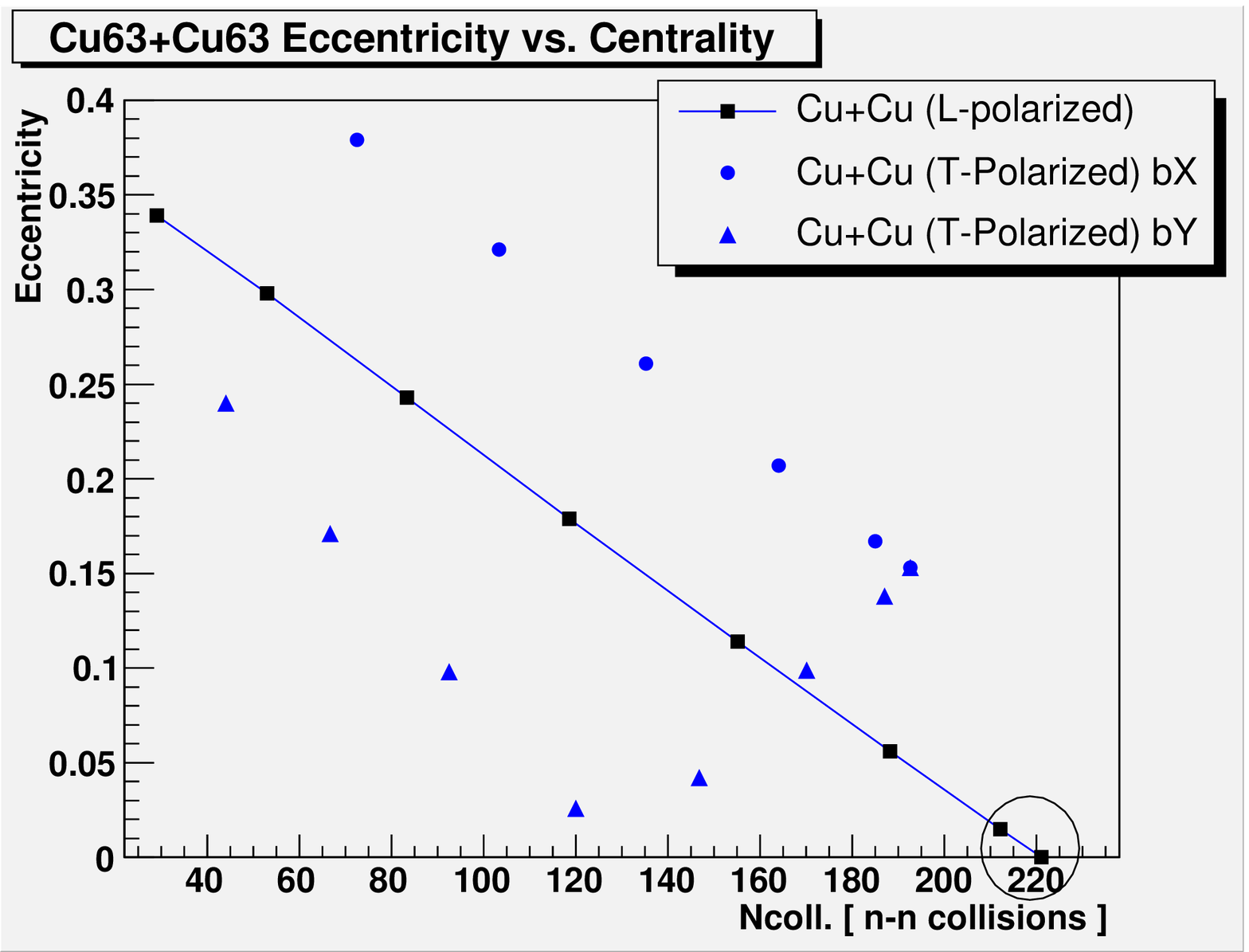}
\includegraphics[width=7cm]{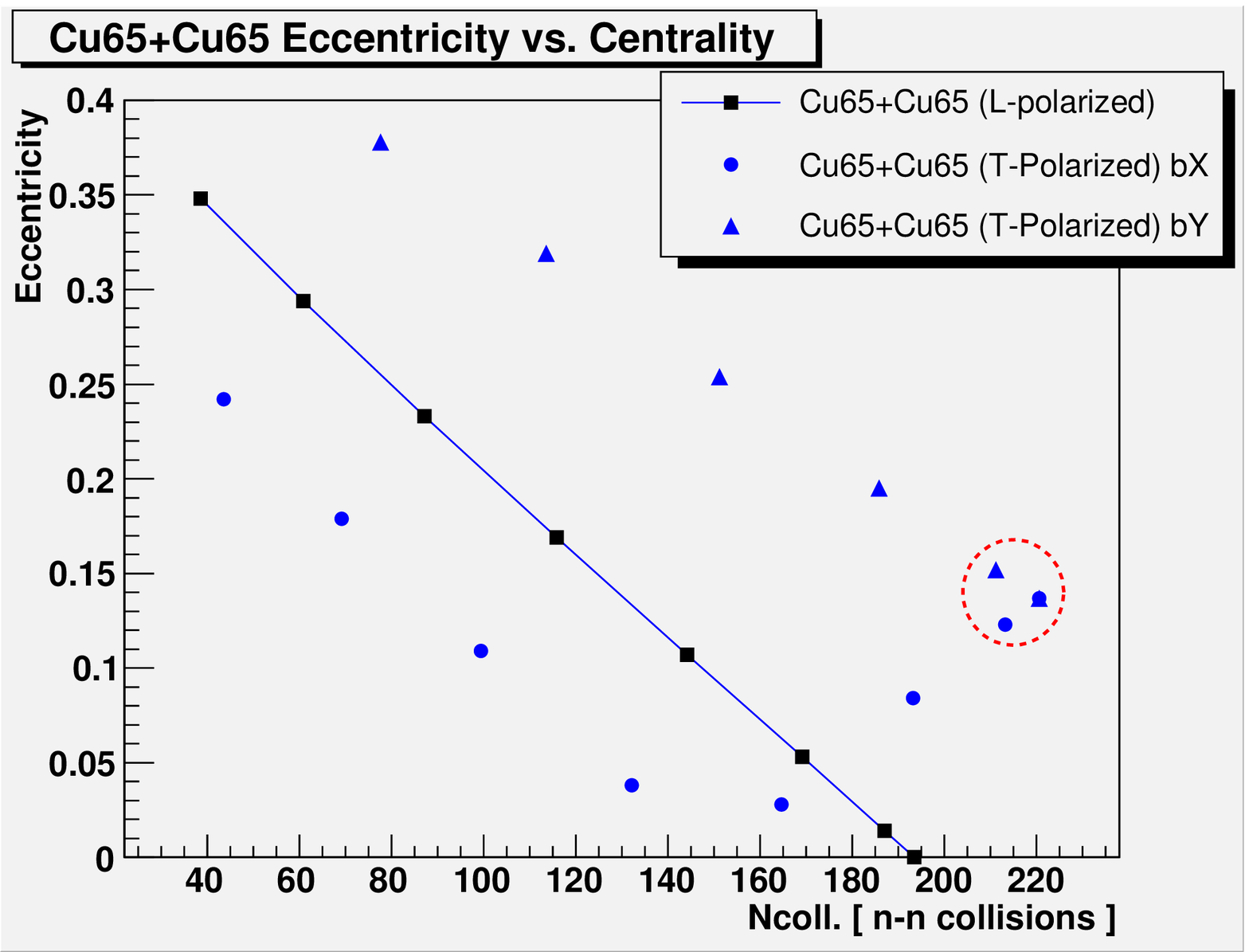}

{\small
{\bf Fig.7:} Ecc[Ncoll] plot for Cu$^{63}$+Cu$^{63}$
and Cu$^{65}$+Cu$^{65}$ assuming $\beta_2 = $ 0.16 and -0.15 \cite{MollerB2}.
}
\label{fig:Shapes7}
\end{figure}

If stable copper isotopes are deformed as
predicted in \cite{MollerB2} then most central collisions of
Cu$^{63}$ should exhibit significantly 
different elliptic flow strength $v_2$
compared to most central collisions of Cu$^{65}$+Cu$^{65}$ 
(predicted $\beta_2$ = -0.15). 
This means that elliptic flow measured in most central collisions
of a given isotope can be used to determine $\beta_2$ sign of its 
ground-state deformation. This might hold also for spin-zero nuclei
e.g. Si$^{28}_{0+}$ (predicted $\beta_2 = -0.48$ \cite{MollerB2}).

Situation is however slightly more complex than indicated in
Figs. 6 and 7. In the case of Trans-Trans collisions
($\theta_1 \approx \theta_2 \approx \pi/2$)
of deformed nuclei there is a freedom in relative azimuthal 
orientation $\Delta \phi$ of nuclear spins. 
For $\Delta \phi \approx \pi/2$ (orthogonal spins) elliptic flow
should tend zero and for $\Delta \phi \approx 0$ (parallel spins) participant
eccentricity and thus also the observed elliptic flow strength $v_2$
should be maximal (whatever the sign of $\beta _2$ is).

\begin{figure}[h]
\includegraphics[width=11cm]{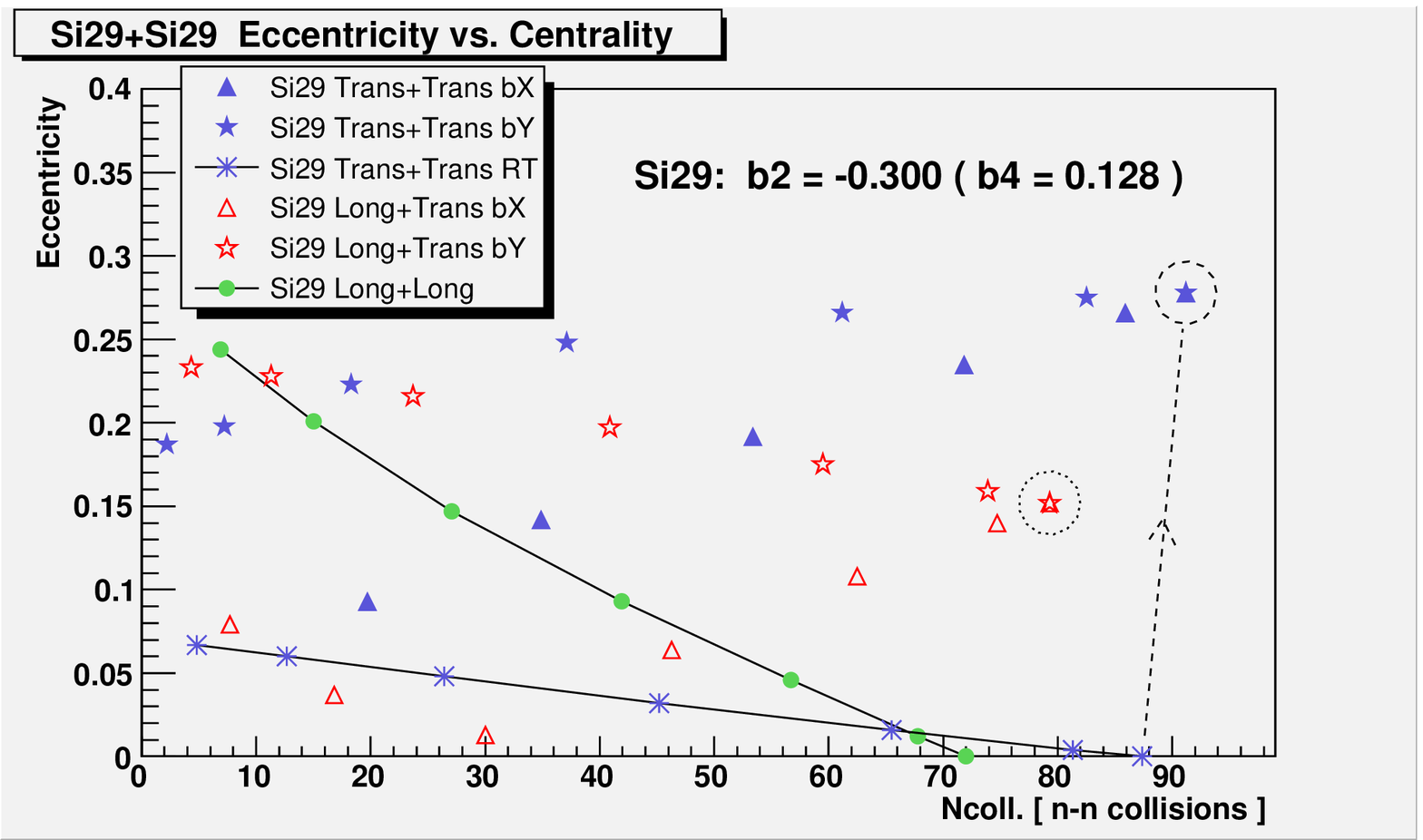}

{\small
{\bf Fig.8:} Eccentricity[Ncoll] plot for 
$^\uparrow$Si$^{29}$+$^\uparrow$Si$^{29}$
assuming $\beta_2$=-0.3 and $\beta_4$=0.13 \cite{MollerB2}.
}
\label{fig:Shapes8}
\end{figure}

This is clearly visible from our calculations in Figure 8, where 
dashed arrow indicates increase of eccentricity
for transversally polarized $\theta_1 = \theta_2 = \pi/2$ 
collisions of Si$^{29}$+Si$^{29}$ when azimuthal angle difference of spins 
changes from $\Delta \phi =\pi/2$ (denoted as Trans-Trans RT)
to parallel spins orientation $\Delta \phi = 0$. Therefore in geometrically most
central collisions (zero impact parameters $b$)
the participant eccentricity
depends on polar angles $\theta_1,\theta_2$ and on relative azimuthal angle 
difference $\Delta \phi$:

\begin{equation}
\epsilon _p [b\approx 0] = \epsilon_p (\theta_1,\theta_2,\Delta \phi)
\end{equation}

For random relative azimuthal angles $\Delta \phi$ a sample of very high multiplicity events 
will contain mixture of collisions with different $\Delta \phi$ and
various eccentricities. This applies also to Au+Au collisions studied at RHIC.
Collisions of longitudinaly+transversaly
polarized Si$^{29}$ have lower maximal number of binary collisions $N_{coll} \approx 80$
(at $b$=0 fm) with eccentricity 
$\tilde \epsilon_{NN} \approx 0.15$ (see Fig.8). Data points in all our Eccentricity[Ncoll.] plots  
correspond to impact parameters $b_x$ and $b_y$
increasing in steps 0,1,2,3,4,5,6 fm. Calculated 
number of binary collsions $N_{coll.}$  
decreases with increasing 
impact parameters ($b_x,0$) and ($0,b_y$).

As a matter of completeness we mention here also the possibility to collide
prolate nuclei ($\beta_2 > 0$) with oblate nuclei ($\beta_2 < 0$)
such as Ho$^{165}$+As$^{75}$ or Cf$^{\,251}$+Au$^{197}$. 
In this case collisions with highest multiplicities
will be Long-Trans polarized and they should exhibit elliptic
flow strength $<v_2>$ rising with $\beta _2$ deformation 
of the prolate nucleus. 

\section{Discussion of results}
It has been shown that in collisions of transversaly polarized prolate
nuclei (e.g. Ho-165) with spherical projectiles (e.g. Pb-207) 
a significat elliptic flow will be generated 
even for zero impact parameters.
Such collisions however will have similar multiplicity
(centrality) as non-central collisions with longitudinaly polarized 
prolate nuclei. 
Careful polarization-dependent studies and precise measurement 
of spectator energy can distinguish different orientations
of prolate nuclei relative to beam axis in such 
collisions. 
Highest-multiplicity collisions of prolate nuclei 
(such as Ho-165) 
correspond to longitudial orientations of prolate nuclei.
Elliptic flow strength $v_2$ will thus tend to zero 
(within fluctuations) in such collisions.

For oblate nuclei the situation is different. Highest multiplicity collisions
of oblate nucleus with a spherical projectile
correspond to transversal polarization of oblate nucleus relative to beam
axis. Additionally if both nuclei colliding are prolate highest multiplicity
collisions will tend to be collisions of transversally polarized nuclei
with parallel azimuthal orientation 
of nuclear deformation axes.
Average value of elliptic flow
strength $v_2$ measured in very high multiplicity (VHM) 
collisions of oblate nuclei
will thus contain two contributions:
\begin{equation}
<v_2>^{VHM} = <v_2>_{fluct} + <v_2>_{\beta_2}
\label{eqX}
\end{equation} 
The first contribution $<v_2>_{fluct.}$ corresponds to eccentricity
fluctuations which are present also in the highest multiplicity collisions 
of spherical nuclei. The second contribution is proportional
to geometrical deformation of oblate ($\beta_2 < 0$) nucleus.
Quantities $<v_2>_{fluct}$ and $<v_2>_{\beta_2}$ can in principle 
be disentangled if elliptic
flow for highest multiplicity collisions of spherical 
nuclei with similar multiplicity
is measured
$<v_2>^{VHM} =  <v_2>_{fluct}$. This approach can be attempted
for two isotopes of a given element (e.g. Si$^{29}$ and Si$^{30}$) 
or for two nuclei
(spherical+oblate) with similar number of nucleons e.g. Pb and Au (see Fig.6).

One should keep in mind still that $<v_2>_{\beta_2}$
contains averaged values of $v_2(\Delta \phi)$ 
due to almost\footnote{The influence of $\Delta \phi$ on collision centrality
at $b\approx 0$ is rather small (see Fig.8 for Si+Si collisions).} 
random relative azimuthal orientations $\Delta \phi$ of nuclear spins 
in such highest multiplicity events (for unpolarized beams and targets).

\section{Conclusions}
A simple simulation based on optical limit approximation of the Glauber model
\cite{Glauber}
shows that very high multiplicity collisions of oblate nuclei at high energies
should exhibit non-zero elliptic flow strength dependent on $\beta _2$
deformation parameter of the oblate nucleus. 
This provides us with the possibility to study the elliptic flow phenomenon
in the most central ultra-relativistic collisions of oblate nuclei.
The equation of state of QCD matter can thus be investigated 
in the highest multiplicity collisions. 
The energy dependence \cite{QGPv1}
of elliptic flow strength $v_2$ in these very high multiplicity
(VHM) central collisions of oblate nuclei can reveal 
changes in the equation state of hadronic matter (e.g. kink-like signatures
\cite{SorgeKink,myKink}) due to the expected phase transition 
of QCD matter. 

For a precise investigation to what extent the oblateness of Au$^{197}$
nucleus influences interpretations drawn from elliptic flow measurements
at RHIC \cite{RHICv2} a full MC simulation based on Glauber model 
taking into account ground-state deformation of Au$^{197}$ nucleus together with
all possible orientations $\theta_1,\theta_2,\phi_1,\phi_2$ of colliding
nuclei
is probably necessary. A comparison of experimental results from Au+Au collisions
with collisions of spherical nuclei 
(e.g. Pb$^{207}$) at the same energy range and the study of Cu+Cu collisions
with the second stable copper isotope at RHIC could be very useful. Elliptic flow
studies in collisions of deformed nuclei at LHC would provide us with new 
and possibly exciting results.

\section{Appendix: Comparison with MC Glauber calculations.}

For a comparison with results of full Monte-Carlo Glauber calculation 
\cite{AuAuRHO}
we present here impact parameter dependencies of some quantities
obtained with our simple optical Glauber model (OGM) assuming
Au$^{197}$ nucleus to be spherical 
($R=6.38$fm; $a=0.53$).

\begin{figure}[h]
\includegraphics[width=9.9cm]{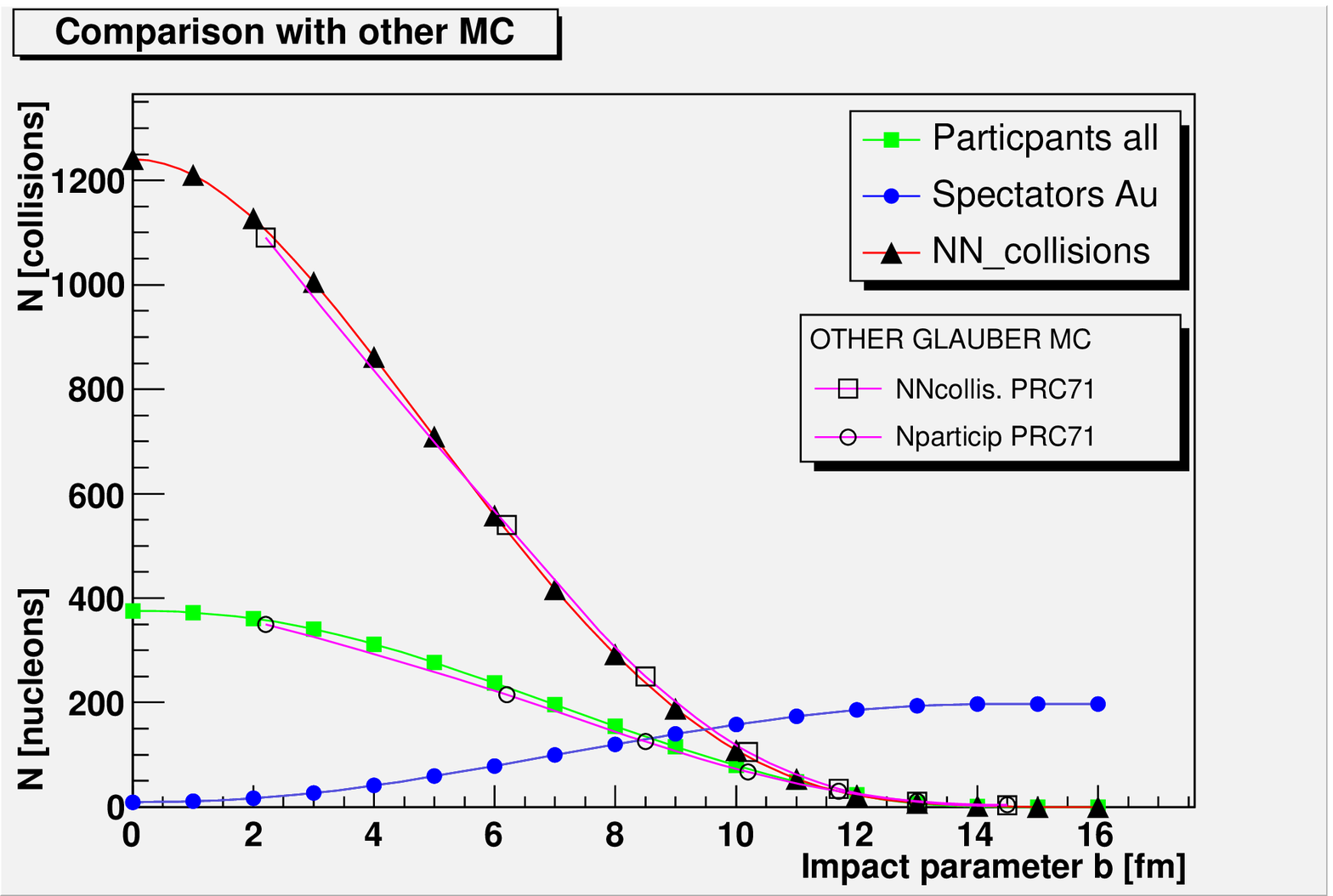}

{\small
{\bf Fig.9:} Centrality dependence
of binary nucleon-nucleon collisions and participants in Au+Au.
}
\label{fig:ShapesC2}
\end{figure}

The impact parameter dependence of 
the number of binary nucleon-nucleon collisions together
with number of participants and spectators calculated in our siplified OGM
model is shown in Figure 9. Data points
from full MC Glauber simulation \cite{AuAuRHO} are shown for a comparison.

In Figure 10 we show the maximal transverse baryon 
density in the overlapping zone
and maximal number of nucleon-nucleon collisions per fm$^2$ in transversal plane
for spherical Au collisions.
Our simple OGM calculations are in reasonable agreement with results
of full MC Glauber simulation \cite{AuAuRHO}.

\begin{figure}[h]
\includegraphics[width=9.9cm]{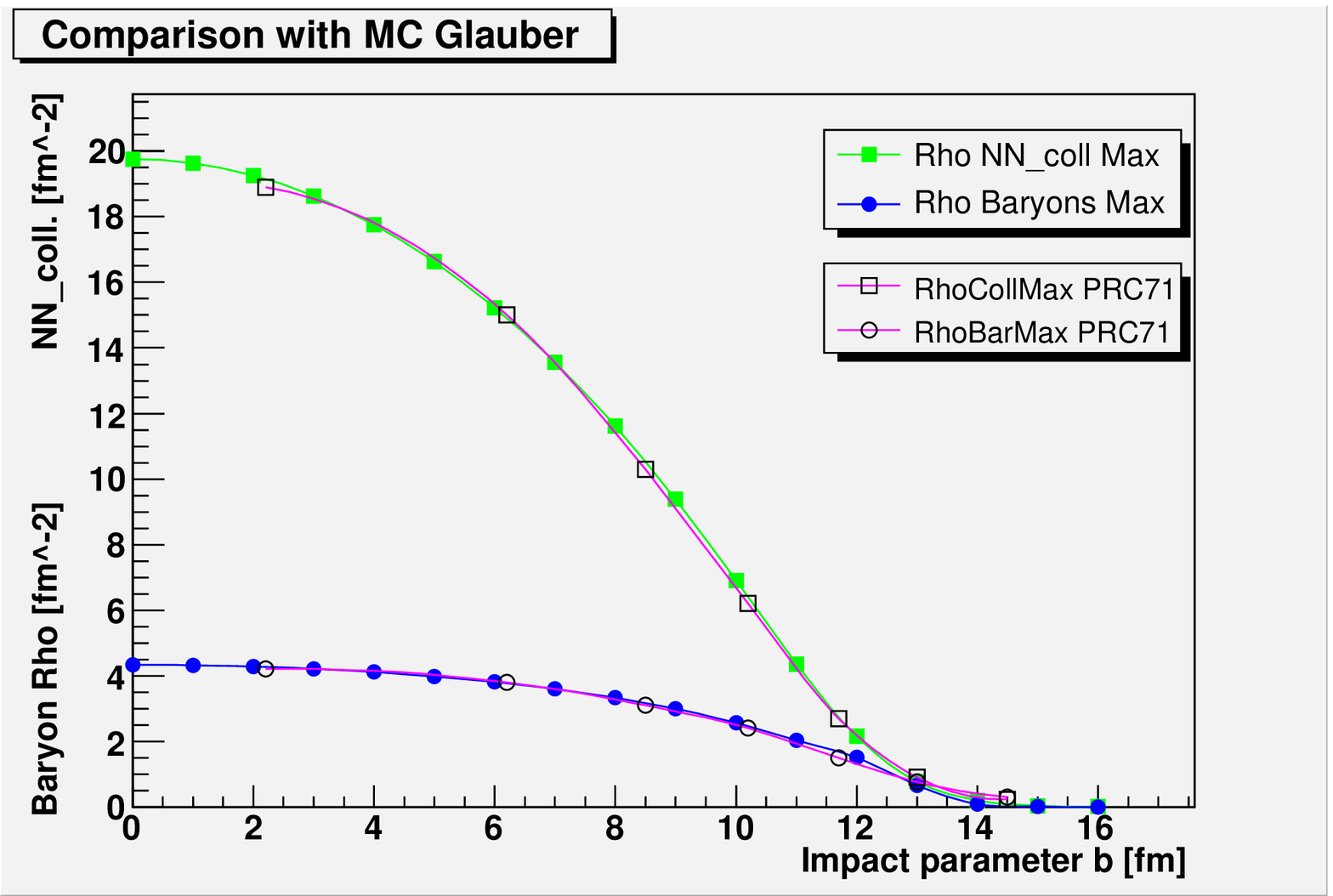} %

{\small
{\bf Fig.10:} Maximal baryon density $\rho_B^{\text{Max}}[b]$ and
maximal collision density $\rho_{\text Ncoll.}^{\text{Max}}[b]$ in Au+Au.
}
\label{fig:ShapesC3}
\end{figure}

Impact parameter dependence of our calculated (OGM) eccentricity $\tilde \epsilon _{NN}$ 
for Au+Au collisions (assuming spherical Au$^{197}$ nucleus) together with eccentricity
values calculated by R.S.Bhalerao et.al. \cite{JYOPLB} is shown in Figure 11. 

\begin{figure}[h]
\includegraphics[width=8.8cm]{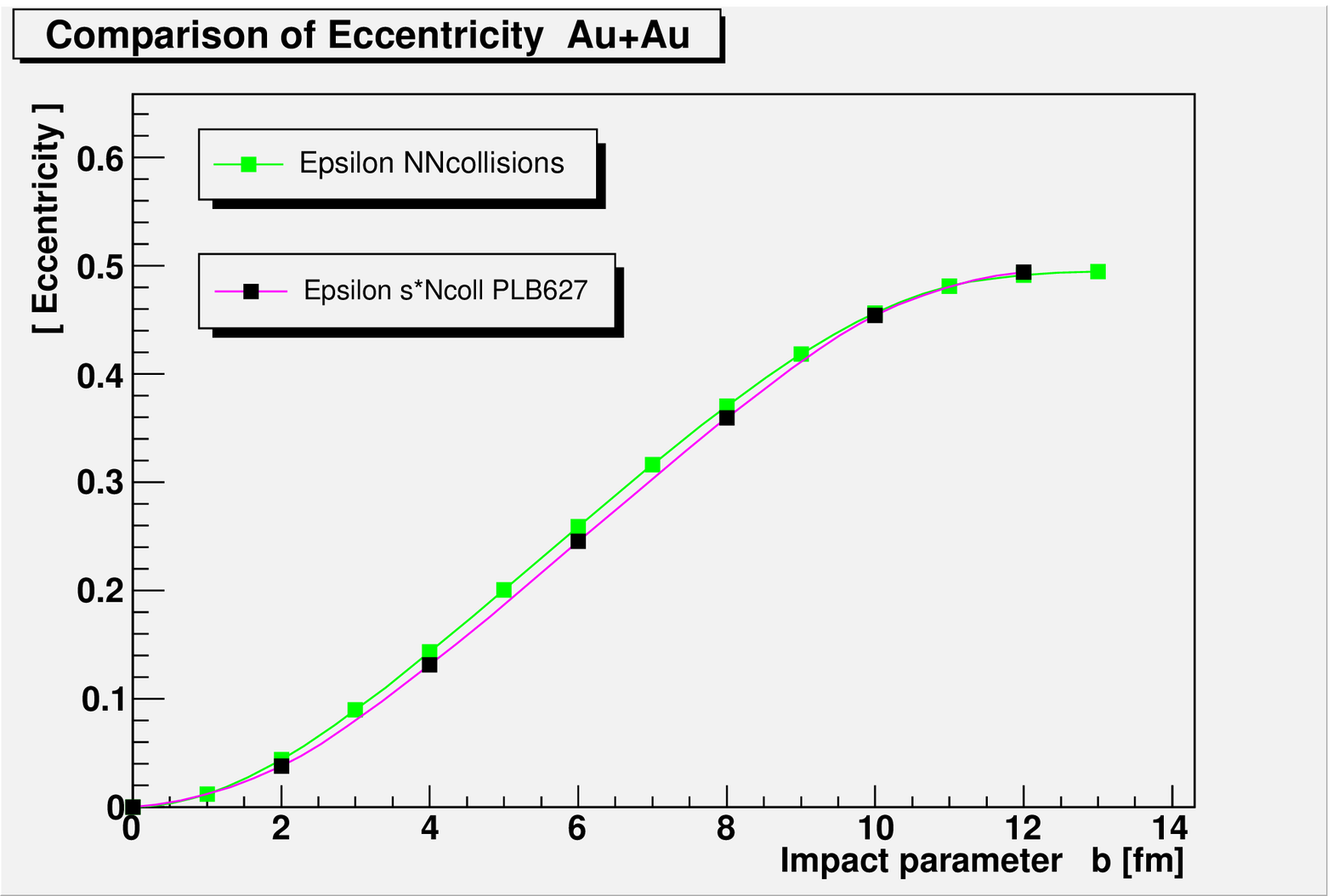}

{\small
{\bf Fig.11:} Impact parameter dependence of (OGM) eccentricity $\tilde \epsilon _{NN}$ 
in comparison with results \cite{JYOPLB}.
}
\label{fig:ShapesC4}
\end{figure}

\newpage
This work was supported by Slovak Grant Agency for Sciences VEGA under
grant N. 2/7116/27.

\end{document}